\documentclass[preprint, onecolumn, superscriptaddress]{revtex4-1}
\usepackage{color}
\usepackage{amsmath,amssymb,graphicx,bm,float}

\newcommand{\ket}[1]{|{#1}\rangle}

\begin{document}
\title{Quantum-memory-assisted multi-photon generation for efficient quantum information processing}
\author{Fumihiro Kaneda} 
\affiliation{Department of Physics, University of Illinois at Urbana-Champaign, Urbara, IL 61801, USA}
\author{Feihu Xu} 
\affiliation{Research Laboratory of Electronics, Massachusetts Institute of Technology, Cambridge, MA
02139, USA }
\author{Joseph Chapman} 
\affiliation{Department of Physics, University of Illinois at Urbana-Champaign, Urbara, IL 61801, USA}
\author{Paul G. Kwiat} 
\affiliation{Department of Physics, University of Illinois at Urbana-Champaign, Urbara, IL 61801, USA}
\begin{abstract}
In the last two decades, many quantum optics experiments have demonstrated small-scale quantum information processing applications with several photons \cite{Pan:2012kv,Tillmann:2013hi,Crespi:2013el}. 
Beyond such proof-of-principle demonstrations, efficient preparation of large, but definite, numbers of photons is of great importance for further scaling up and speeding up photonic quantum information processing \cite{Kok:2007ep,Aaronson:2011ja,Childs:2013hh}. 
Typical single-photon generation techniques based on nonlinear parametric processes face challenges of probabilistic generation. 
Here we demonstrate efficient synchronization of photons from multiple nonlinear parametric heralded single-photon sources (HSPSs), using quantum memories (QMs). 
Our low-loss optical memories greatly enhance ($\sim 30\times$) the generation rate of coincidence photons from two independent HSPSs, while maintaining high indistinguishability ($95.7 \pm 1.4$\%) of the synchronized photons. 
As an application, we perform the first demonstration of HSPS-based measurement-device-independent quantum key distribution (MDI-QKD).
The synchronized HSPSs demonstrated here will pave the way toward efficient quantum communication and larger scale optical quantum computing. 
\end{abstract}

\maketitle

For single- and multi-photon generation, quantum optics experiments have typically used nonlinear optical parametric sources due to experimental convenience and their stable performances (in contrast, solid-state single-emitter sources \cite{Aharonovich:2016hm} typically require cryogenic cooling and suffer from source inhomogeneity and difficulty achieving high-efficiency collection of emitted photons into a single spatial mode, e.g., optical fiber). 
However, these cannot generate single photon pairs deterministically; 
for a mean number of photon pairs $\mu$, the generation probability of $k$ photon pairs is $\mu^k / (\mu +1 )^{k+1}$. 
Therefore, the single-pair generation probability peaks at only 25\% due to the non-negligible likelihood ($\sim \mu^k$) of unwanted zero- and multiple-pair generations. 
For example, a recent ten-photon experiment \cite{Wang:2016dk} using 5 spontaneous parametric downconversion (SPDC) sources needed to keep $\mu < 0.05$ to suppress the multi-pair emissions, resulting in a ten-photon coincidence rate of only several events per hour.

Here we employ QMs to synchronize such probabilistic parametric sources to efficiently generate multiple simultaneous single photons, as shown in Fig. \ref{Idea}. 
$M$ parametric sources pumped with a period $\tau$, generate photon pairs probabilistically, though in general not simultaneously. 
Each parametric source works as a HSPS in which photons generated in a trigger mode are sent to a single-photon detector (SPD), whose click ``heralds'' in which time slot the corresponding twin photon is present. 
Each QM triggered by a heralding signal from its corresponding HSPS, stores heralded photons for an arbitrary integer time of $\tau$, until other sources produce their pairs.  
After the last source heralds a ``last-born'' photon, the $M-1$ memories storing the earlier-born photons release them simultaneously, 
thereby producing $M$ simultaneous photons. 
Given each source's heralding probability per pump pulse $p \sim \mu \eta \ll 1$ (where $\eta$ is the system detection efficiency of the trigger mode), a maximum number of storage time slots $N$, and lossless QMs, the $M$-fold coincidence probability is given by $\{1-(1 - p)^N \}^M \simeq (p N)^M$. 
Hence, one can obtain up to $\times N^{M-1} $ enhancement over a non-synchronized case that requires $M$ sources to simultaneously herald $M$ photons (with probability $p^M N$). 
Theoretical details of the synchronization scheme are discussed in Ref. \cite{Nunn:2013hua, GimenoSegovia:2017vl} and Supplementary Information. 
Similar schemes have been demonstrated by using optical parametric oscillators \cite{Makino:2016bi} and atomic ensembles \cite{Felinto:2006gw}; however, our pulsed-pump scheme is advanced in high-speed capability.  
Note that this synchronization scheme even has a higher generation rate compared to recently demonstrated \textit{periodic} time-multiplexed HSPSs \cite{Kaneda:2015dn,Mendoza:2016gr,Xiong:2016bv}: 
$M$ periodic time-multiplexed sources need to wait for periodic output time windows even if all QMs have loaded photons earlier.  
In contrast, our proposed scheme needs to store $M-1$ photons only for the \textit{difference} of the generation time slots, substantially reducing total storage loss in imperfect (and practical) QMs. 
Also, the synchronization process can be repeated immediately after the last source heralds its photon.

\begin{tiny}
\begin{figure}[t!]
  \includegraphics[width=0.5\columnwidth,clip]{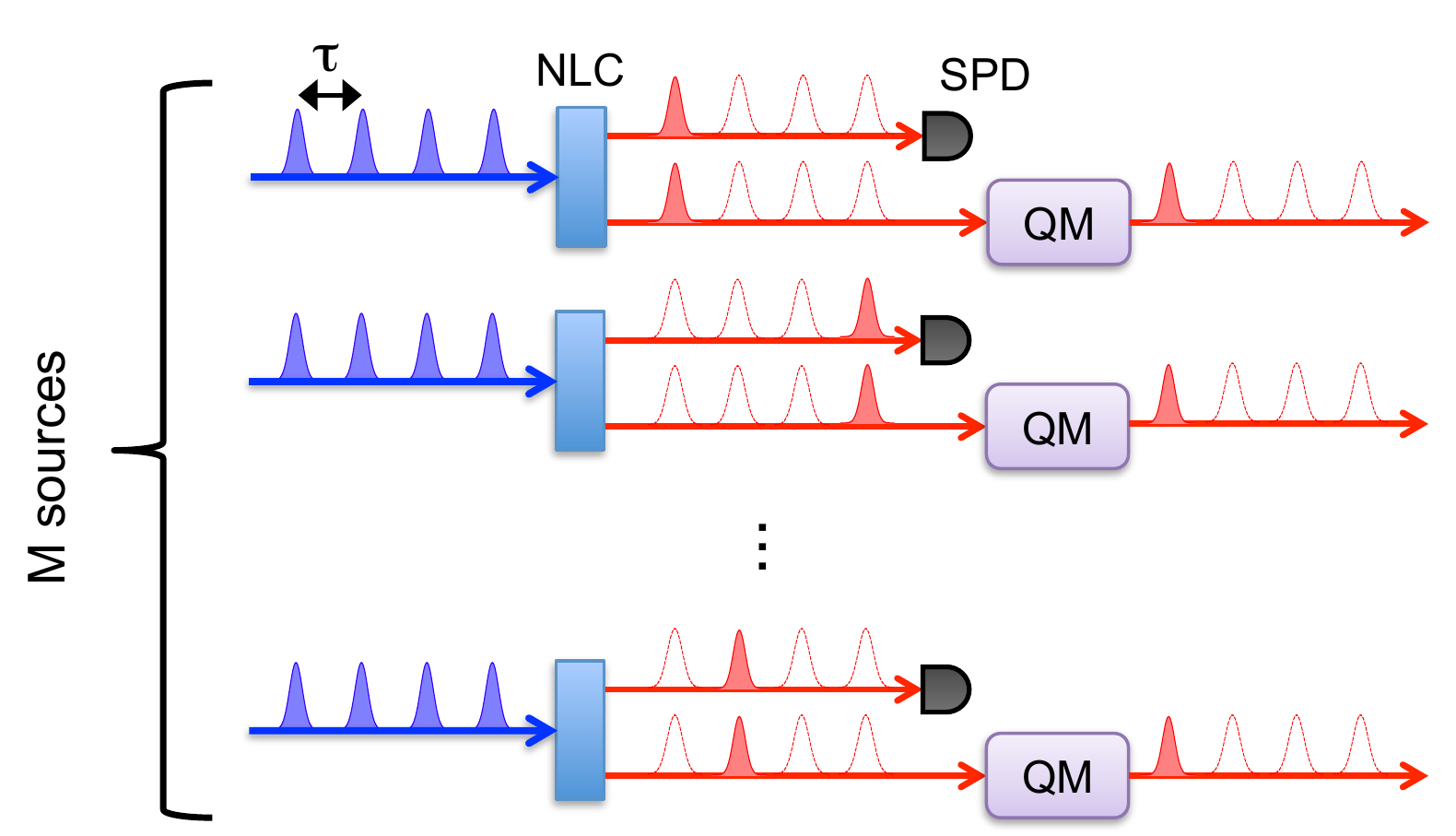}
   \caption{A scheme to generate $M$ single photons from $M$ HSPSs synchronized by quantum memories. NLC, nonlinear crystal; SPD, single-photon detector; QM, quantum memory. NLCs in general produce single photon pairs only rarely, and thus simultaneous $M$-photon generation occurs only with very low probability. QMs can compensate for the relative delay of photons from each source, and release them simultaneously.}
\label{Idea}
\end{figure}
\end{tiny}

\begin{tiny}
\begin{figure}[t!]
  \includegraphics[width=0.79\columnwidth,clip]{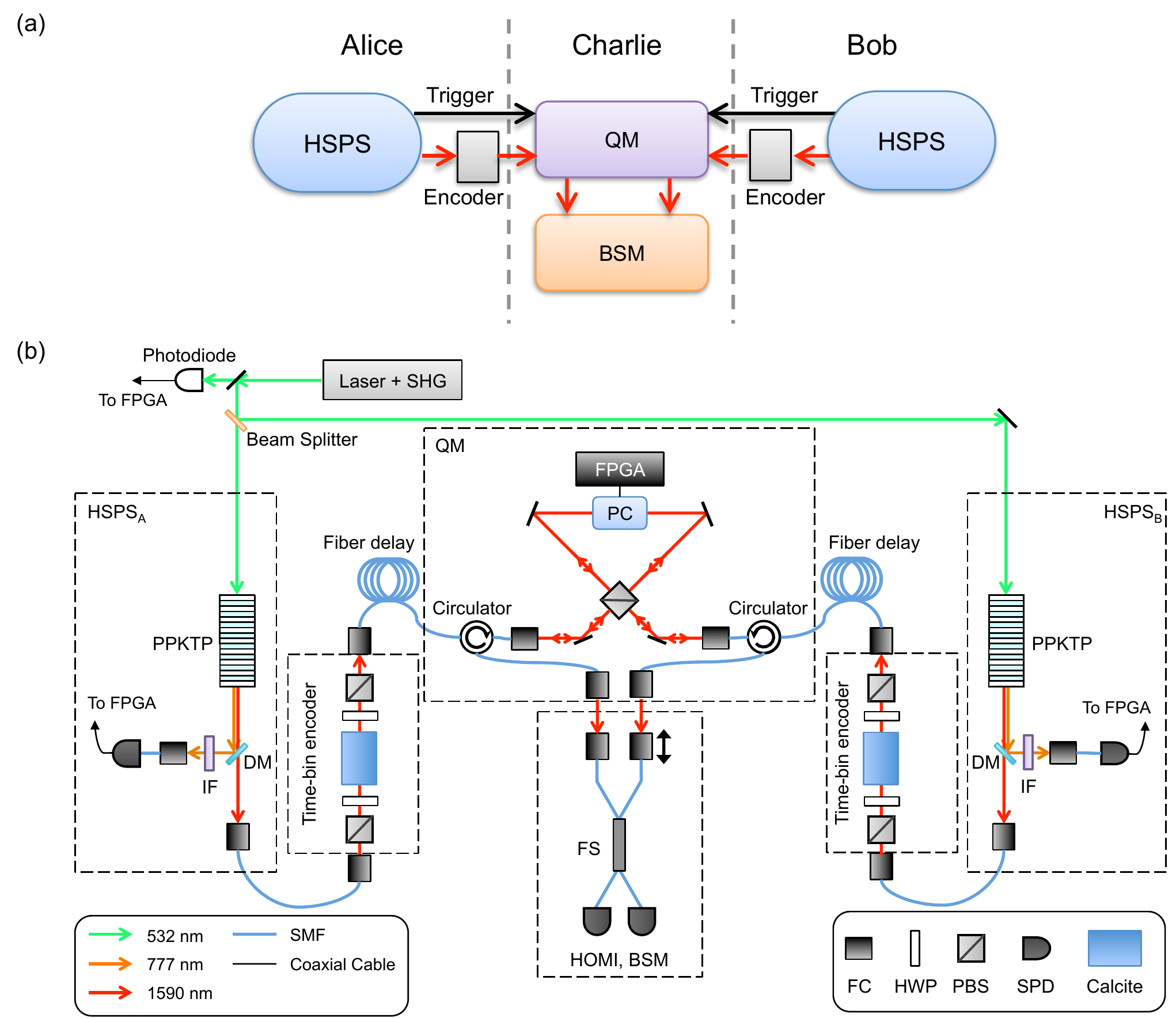}
   \caption{(a) Our proposed MDI-QKD scheme, in which Charlie can synchronize the photons from two remote HSPSs. 
   In MDI-QKD, Alice and Bob, each possess qubit encoders and probabilistic single-photon sources, e.g., HSPSs or faint laser sources. 
Charlie receives Alice's and Bob's photons, performing a Bell-state measurement (BSM) on them. 
Informed by Charlie which Bell state he observed, Alice and Bob know the specific correlation between their respective qubits, perform post-processing and generate a shared secure key. 
Since MDI-QKD requires two-photon coincidences in the BSM, efficient simultaneous generation of single-photon states is more critical to realize high secure key rate, while a traditional BB84 protocol in principle needs only one single-photon source (but then requires additional assumptions about the detectors \cite{Lo:2014ex}). 
A QM module in our scheme delays an early-arrival photon to be sent to the BSM setup simultaneously with a late-arrival photon. (b) Schematic diagram of our experimental setup, with FC, fiber coupler; HWP, half-wave plate; PBS, polarizing beam splitter; SPD, single-photon detector; PC, Pockels cell; IF, interference filter ($\Delta \lambda$ = 1.1 nm);  DM, dichroic mirror; SMF, single-mode fiber; and FPGA, field-programmable gate array. See Methods for experimental details. }
\label{Idea2}
\end{figure}
\end{tiny}

Our scheme in general can be applied to multiple HSPSs not only in a local laboratory together but also in remote locations; the former case is very useful for quantum computing applications \cite{Kok:2007ep,Aaronson:2011ja,Childs:2013hh}, while the latter has great potential for realizing efficient quantum networking. 
Particularly, synchronized remote sources can be directly applicable to an important quantum communication application, MDI-QKD \cite{Lo:2012df}, that is secure against all detector side-channel attacks.
Our proposed MDI-QKD scheme with QMs is depicted in Fig. \ref{Idea2} (a). 
In general MDI-QKD, Alice and Bob, who want to share secure cryptographic keys with each other, both need to simultaneously send qubit-encoded photons to Charlie, who identifies the correlation between Alice's and Bob's qubits (but not those qubits themselves) via Bell-state measurement (BSM), i.e., projection measurement of them into the Bell-state basis. 
Therefore, in MDI-QKD efficient simultaneous generation of single-photon states is more critical to produce higher secure key rates than a traditional BB84 protocol \cite{Lo:2014ex}. 
In our scheme, in addition to the standard BSM configuration, Charlie possesses a QM module so that an early-arrival photon from Alice's (Bob's) HSPS is delayed to be sent to the BSM setup simultaneously with a late-arrival photon from Bob's (Alice's) source. 
Thus, the success event rate of the BSM and thereby the secure key rate and transmission distance are significantly increased compared to the standard (non-synchronized) case \cite{Abruzzo:2014it,Panayi:2014jr}.

A schematic diagram of our experiment for synchronizing two HSPSs is shown in Fig. \ref{Idea2} (b). 
Our HSPSs \cite{Kaneda:2016fh} pumped by a common pulsed laser source (with period $\tau = 10$ ns) generate heralded photons at 1590 nm with a 96\% spectral indistinguishability. 
We operated the pair generation rate at $\mu=0.013$ per pulse, for which the SPDC multi-pair contribution to the total coincidence counts was limited to $\sim$4\%.
Our QM, consisting of a bulk optics delay cavity with a matched cycle length $\tau$ and a high-speed polarization switch (a Pockels cell, PC), has 98.8\% transmission per cycle. 
Incorporating two fiber optic circulators, the QM can delay photons coming from \textit{either} of the HSPSs (see Methods). 
Due to the low switching rate (1 MHz) of the PC, the synchronization process is not repeated immediately after synchronizing two photons, but after a fixed cycle (every 1 $\mu$s). 

Figure \ref{result1} (a,b) respectively show synchronized trigger signal rates from two HSPSs and two-photon coincidence count rates versus $N$. 
The synchronized trigger signal rate increases approximately as $N^2$ as expected; 
an $\sim N^2$ increase is also observed for the two-photon coincidence count rates, due to the high storage efficiency. 
Without the synchronization process, we observed a coincidence count rate of only $121 \pm 6$ per 100 s with the pump repetition rate of $1/\tau = 100$ MHz. 
We determined the enhancement factor for the two-photon coincidence count rate as the ratio of the synchronized and non-synchronized case's coincidence count rates per pump pulse (see Fig. \ref{result1} (c)); 
the enhancement factor increases almost linearly as $N$, and $\times 30.5 \pm 1.6$ enhancement was obtained with $N= 40$. 
Note that this same approach, generalized to preparing, e.g., 10 simultaneous photons, would have an enhancement factor of $30.5^9 = 2.28\times10^{13}$. 
Our results are in agreement with the theoretical predictions, shown as solid lines in Fig. \ref{result1} (a,b,c) (see Supplementary Information).

\begin{tiny}
\begin{figure}[t!]
\begin{center}
  \includegraphics[width=0.78\columnwidth,clip]{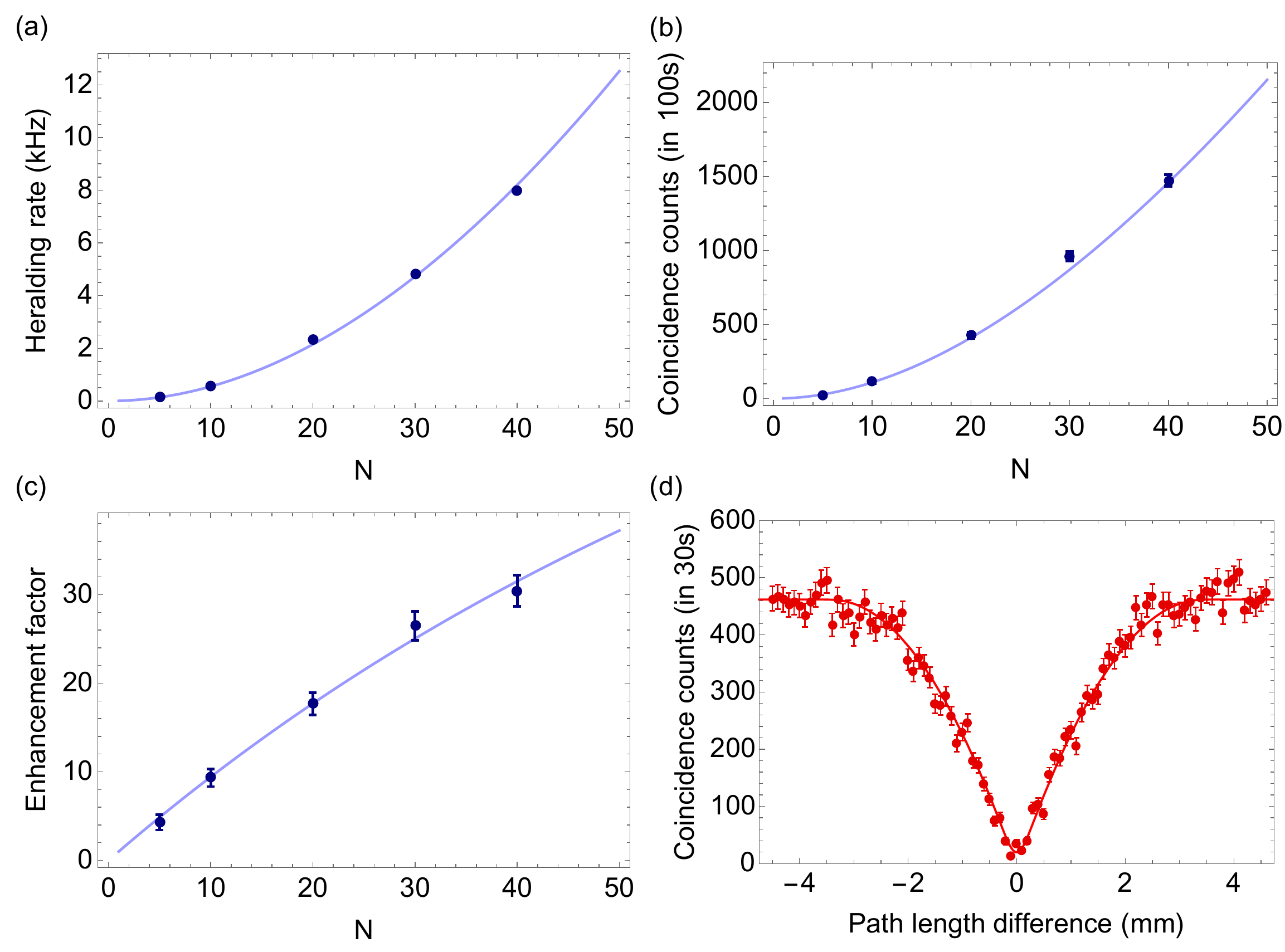}
   \caption{(a) Synchronized trigger signal rate, (b) coincidence count rate of synchronized heralded photons, and (c) enhancement factor of coincidence count rates in the synchronized case compared to the asynchronous case, versus maximum storage cycle number $N$. (d) Observed HOMI for synchronized photons, with $N = 40$. Error bars are estimated by Poissonian photon counting statistics.}
\label{result1}
\end{center}
\end{figure}
\end{tiny}

We characterized the indistinguishability of the synchronized photons by Hong-Ou-Mandel interference (HOMI) \cite{Hong:1987vi}, of which visibility is a direct measure, and essential for BSM (as will be demonstrated). 
Our observed HOMI with $N = 40$ as well as the best-fit theoretical curve \cite{Kaneda:2016fh}, is shown in Fig. \ref{result1} (d). 
The estimated visibility and dip width after subtracting background counts (23.2 counts for each data point) were $95.7\pm 1.5$\% and $6.00\pm 0.02$ ps, respectively, which closely matches our prediction based on the observed joint spectral intensities of the HSPSs (see Supplementary Information); 
the background counts are mainly due to the multi-photon emissions.  
This high HOMI visibility indicates that our QM well preserves the time-bandwidth characteristics ($\Delta t = 6.1$ ps, $\Delta \lambda = 0.8$ nm) and indistinguishability of the heralded photons.

\begin{tiny}
\begin{figure}[t!]
  \includegraphics[width=0.78\columnwidth,clip]{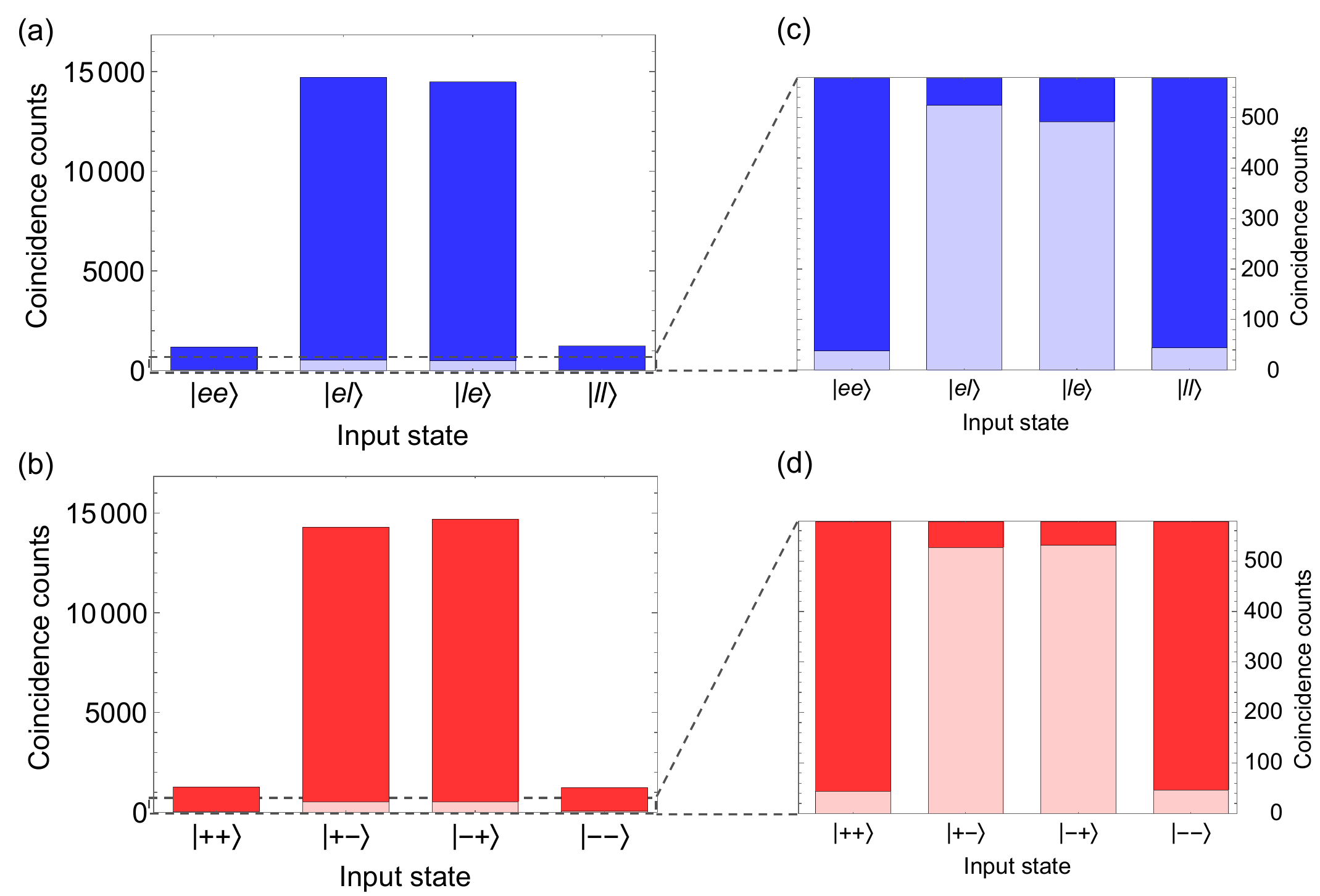}
   \caption{Coincidence counts (without subtracting background counts) from BSM of different time-bin encoded photons. (a)  $\{\ket{e}, \ket{l}\}$ basis. (b) $\{\ket{+}, \ket{-} \}$ basis. We observed very high contrast in coincidence count rates, implying low quantum bit error rates. (c,d) Enlarged figures of (a,b), showing the corresponding coincidence counts without synchronization, in light colors; while error probabilities are similar, the count rates are $\sim 30\times$ lower.  Coincidence counts are collected for $N_p = 4\times 10^9$ pump pulses in each data point. 
}
\label{result2}
\end{figure}
\end{tiny}

Lastly, we apply our synchronization technique to demonstrate proof-of-concept MDI-QKD with time-bin-encoded heralded single photons.  
Note that polarization qubits are not switchable, because our QM switches polarization to control a photon's delay, so instead we use time-bin encoding  (see Methods).
Figure \ref{result2} (a,b) shows experimental results of the BSM for the early-/late-qubit basis $\{\ket{e}, \ket{l}\}$ and their superposition basis $\{\ket{+}, \ket{-} \}$, where $\ket{e} \perp \ket{l}$, $\ket{\pm} = (\ket{e} \pm \ket{l})/\sqrt{2}$. 
Coincidence events are collected for $N_p = 4\times 10^9$ pump pulses. 
With our BSM setup projecting two qubits onto a singlet state $\ket{\psi^-} \equiv (\ket{el}-\ket{le})/\sqrt{2} = (\ket{+-}-\ket{-+})/\sqrt{2}$ (see Methods), 
our observed coincidence counts (without subtracting background counts) from identical qubits are only $\sim$ 8\% of those from orthogonal qubits, due to high-visibility HOMI ($\sim$92\%). 
Note that these highly suppressed error count rates depend on the low multi-photon contributions ($\sim$4\%); in contrast, previous demonstrations \cite{Rubenok:2013ip,PhysRevA.88.052303,PhysRevLett.112.190503,Comandar:2016em,PhysRevLett.117.190501} of MDI-QKD with weak coherent pulses could only have 50\% HOMI visibility because of their large photon-number noise.

Based on the result of the BSM, we estimate the lower bound of secure key rate $R = 0.212 \times 10^{-7}$ bit per pump pulse (corresponding to 0.851 bit/s with our 1-MHz system repetition rate) over an equivalent loss, i.e., the total loss of two optical channels from each SPDC crystal to the first circulator, of $\sim$14 dB. 
See Supplementary Information for details of secure key rate evaluations. 
For comparison, we also performed our MDI-QKD experiment without synchronization. 
Although a similar distribution of coincidence counts is observed (see Fig. \ref{result2} (c,d)), no positive key could be guaranteed because of the large uncertainty in the estimates of the QKD bit error rates, due to $\sim 30\times$ fewer photon count rates compared to the synchronized case. 
Therefore, the enhanced coincidence count rate with our synchronization technique is critical to enable useful HSPS-based MDI-QKD. 

Our current secure key generation rate could be enhanced by a factor of $\sim 250$ by several improvements on our current physical setup (see Supplementary Information). 
In addition, we expect that employing decoy-state methods would allow us to use much higher values of $\mu$,  thereby further increasing the secure key rate \cite{Abruzzo:2014it,Panayi:2014jr}. 
Furthermore, passive decoy-state methods \cite{PhysRevLett.99.180503, PhysRevA.75.050305} can be applied for HSPS-based MDI-QKD to remove active decoy intensity modulations. 

In conclusion, we have demonstrated QM-assisted synchronization of multiple HSPSs for efficiently generating multiple single-photon states. 
Our synchronization scheme can be applied with both local and remote HSPSs; the former is valuable for larger scale quantum computing, while the latter has great potential of realizing efficient and low-noise quantum communication. 
We observed greatly enhanced coincidence count rates as well as high indistinguishability of photons from two synchronized HSPSs. 
Moreover, for the first time we obtained secure keys via HSPS-based MDI-QKD, with the help of the source synchronization.
We anticipate that these synchronization methods will pave the way toward larger scale optical quantum computation and communication applications.

\section*{Methods}
\subsection*{Heralded single-photon source}
We used a frequency-doubled mode-locked Yb laser ($\lambda = 521$ nm, $\tau = 10.0 $ ns) to pump two 20-mm-long periodically-poled potassium titanyl phosphate (PPKTP) crystals each of which generates collinear photon pairs centered at 777 and 1590 nm via SPDC. 
We used a shared pump merely for a convenience, not necessity; it is feasible to have independent but locked pump lasers \cite{Kaltenbaek:2006ch}. 
The spectral purity is estimated to be 97\% after filtering the original 2.5-nm bandwidth of the 777-nm mode with 1.1-nm bandpass filters. 
Due to the spectral filtering, we observed largely different transmissions in the two SPDC modes after a collection SMF; the transmission for the heralded (1590-nm) mode is 88\%, while only  30\% for the trigger (777-nm) mode. 
Each trigger detector, a Si avalanche photodiode, has a $\sim$60\% detection efficiency and $10^{-6}$ background count rate per 1-ns coincidence window. 

\subsection*{Quantum memory and synchronization procedure}
We implemented a bulk-optics-based QM, consisting of a 10-ns delay loop, custom Brewster-angled PBS, and PC comprising a pair of rubidium titanyl phosphate (RTP) crystals. 
The two PBS inputs allow the QM to delay photons from either of the sources, with photons from different HSPSs always cycling in opposite directions in the optical delay cavity. 
A field-programmable gate array (FPGA) module processes input signals from trigger SPDs, triggering the PC to store/release early-born photons. 
When an early-born photon from either source enters into the cavity, the PC is activated, rotating that photon's polarization by 90$^\circ$ to store and delay it in the cavity. 
To switch a photon from one HSPS into and out of the QM without affecting a potential photon from the other HSPS, the two sources have a time-slot offset by $\tau /2 = 5$ ns, greater than the 4-ns rise/fall time of our PC. 
After delaying the early-born photon for the necessary integer multiples of $\tau$, photons from the two HSPS are synchronously emitted (but offset by $\tau/2$) from the different ports of the PBS, each coupling to a fiber circulator that directs it to a fiber splitter (whose input arm lengths are chosen to remove the $\tau/2$ offset between the two photons). 
The single-pass cavity transmission $T_c$ is 98.8\%, and the corresponding photon lifetime in the QM is $\sim$ 830 ns (i.e., 83 cycles for $1/e$ total switching transmission).   
The slightly imperfect cavity transmission is due to the transmission of the PC ($99.2$\%) and the reflection of the two concave mirrors (99.8\%). 
The group-velocity dispersion in the QM cavity is very small ($\sim$$1.2\times 10^{-3}$ $ \mathrm{ps^2}$ at 1590 nm) compared to the photon coherence time ($\Delta t =6.1$ ps); thus, the cycle-dependent chromatic dispersion, which could degrade indistinguishability of the synchronized photons, is negligible for up to $N = 40$. 
Each fiber delay line can hold photons for $\sim$500 ns to compensate for the electronic latencies ($\sim 100$ ns from a trigger photon to firing the PC). 
In addition, the rest of the delay ($> 400$ ns) after the compensation allows us to select the \textit{latest} heralded time slot (for up to $N =40$) of the first-heralding HSPS, thus minimizing the effective storage loss in the QM \cite{Kaneda:2015dn} (for example, if HSPS$_A$ produces photons in time slots 3 and 29, and HSPS$_B$ produces a photon in slot 31, we only need to store the second HSPS$_A$ photon for 2 cycles instead of 28).  

\subsection*{Time-bin encoder}
A time-bin qubit state is created by using a common-path polarization-dependent unbalanced interferometer. 
For this proof-of-concept MDI-QKD experiment, no random number generator or fast active switch is used to encode qubits. 
Horizontally polarized photons generated from an HSPS first pass through a HWP with its optic axis at either 0$^\circ$, 45$^\circ$, $22.5^\circ$, or $-22.5^\circ$,  respectively creating the horizontal, vertical, diagonal, or anti-diagonal state ($\ket{H}$, $\ket{V}$, $\ket{D}$, or $\ket{A}$), where $\ket{H} \perp \ket{V}$, $\ket{D}  \equiv (\ket{H} + \ket{V})\sqrt{2}$, and $\ket{A}  \equiv (\ket{H} - \ket{V})\sqrt{2}$.  
A pair of 40-mm-long calcite crystals provides a group delay ($\sim$ 25 ps) between $\ket{H}$ and $\ket{V}$ without transverse walk-off, correlating the polarization state to a temporal one, i.e., $\ket{H} \rightarrow \ket{H} \ket{e}, \ket{V} \rightarrow \ket{V} \ket{l}$.
This group delay is much larger than $\Delta t$ but much smaller than the 4-ns switching rise/fall time of our PC, so that both time-bin states can be efficiently switched in the PC. 
A HWP at 22.5$^\circ$ after the calcite crystals rotates the polarization from $\ket{H}$ ($\ket{V}$) to $\ket{D}$ ($\ket{A}$), and a following PBS transmits only $\ket{H}$. 
Thus, time-bin qubit states ($\ket{e}$, $\ket{l}$, $\ket{+} \equiv  (\ket{e} + \ket{l})/\sqrt{2}$, $\ket{-} \equiv (\ket{e} - \ket{l})/\sqrt{2}$) with an identical polarization state $\ket{H}$ are successfully generated with a 50\% postselection probability; the overall transmission including optics loss, fiber coupling efficiency, and this postselection, is about 22\%.

\subsection*{Measurement of synchronized photons}
In order to perform the HOMI experiment as well as the BSM, we implemented an interferometer with a 50:50 fiber splitter. 
The path length difference between the two fiber input mode is adjusted to be zero by translating one of input fiber couplers. 
Coincidence counts of the synchronized photons are measured by two fiber-coupled superconducting nanowire detectors (SNSPDs) with $\sim$75\% detection efficiency and $\sim10^{-6}$ dark count probability per 1-ns coincidence window. 
This setup with zero path-length difference and coincidence measurements performs as a BSM projecting onto $\ket{\psi^-} \equiv (\ket{el}-\ket{le})/\sqrt{2} = (\ket{+-}-\ket{-+})/\sqrt{2}$ for time-bin qubits \cite{Rubenok:2013ip,PhysRevLett.111.130502,PhysRevLett.117.190501}. 
For the measurement of coincidence counts versus $N$ shown in Fig. \ref{result1} (b), we used the large path-length difference ($\sim$15 mm) of the interferometer to avoid two-photon interference. 

\section*{Acknowledgement}
Funding for this work has been provided by NSF Grant No. PHY 12-12439 and PHY 15-20991, US Army ARO DURIP Grant No. W911NF-12-1-0562, ARO Grant No. W911NF-13-1-0402, and US Navy ONR MURI Grant No. N00014-13-1-0627.



\begin{thebibliography}{10}
\newcommand{\enquote}[1]{``#1''}

\bibitem{Pan:2012kv}
J.-W. Pan, Z.-B. Chen, C.-Y. Lu, H.~Weinfurter, A.~Zeilinger, and
  M.~{\.Z}ukowski, \enquote{{Multiphoton entanglement and interferometry},}
  Rev. Mod. Phys. \textbf{84}, 777 (2012).

\bibitem{Tillmann:2013hi}
M.~Tillmann, \textit{et al.},
  \enquote{{Experimental boson sampling},} Nature Photon. \textbf{7}, 540
  (2013).

\bibitem{Crespi:2013el}
A.~Crespi, \textit{et al.}, \enquote{{Integrated
  multimode interferometers with arbitrary designs for photonic boson
  sampling},} Nature Photon. \textbf{7}, 545 (2013).

\bibitem{Kok:2007ep}
P.~Kok, K.~Nemoto, T.~C. Ralph, J.~P. Dowling, and G.~J. Milburn,
  \enquote{{Linear optical quantum computing with photonic qubits},} Rev. Mod. Phys. \textbf{79}, 135 (2007).

\bibitem{Aaronson:2011ja}
S.~Aaronson and A.~Arkhipov, \enquote{{The computational complexity of linear
  optics},} in \enquote{Proceedings of the 43rd annual ACM symposium on Theory
  of computing,}  (ACM, New York, 2011), p. 333.

\bibitem{Childs:2013hh}
A.~M. Childs, D.~Gosset, and Z.~Webb, \enquote{{Universal computation by
  multiparticle quantum walk},} Science \textbf{339}, 791 (2013).


\bibitem{Aharonovich:2016hm}
I.~Aharonovich, D.~Englund, and M.~Toth, \enquote{{Solid-state single-photon
  emitters},} Nature Photon.  \textbf{10}, 631 (2016).

\bibitem{Wang:2016dk}
X.-L. Wang, \textit{et al.}, \enquote{{Experimental Ten-Photon
  Entanglement.}} Phys. Rev. Lett. \textbf{117}, 210502 (2016).

\bibitem{Nunn:2013hua}
J.~Nunn, \textit{et al.},
  \enquote{{Enhancing Multiphoton Rates with Quantum Memories},} Phys. Rev. Lett. \textbf{110}, 133601 (2013).

\bibitem{GimenoSegovia:2017vl}
M.~Gimeno-Segovia, \textit{et al.},
  \enquote{{Relative multiplexing for minimizing switching in linear-optical
  quantum computing},} arXiv:1701.03306  (2017).

\bibitem{Makino:2016bi}
K.~Makino, \textit{et al.}, \enquote{{Synchronization of optical photons for quantum
  information processing.}} Sci. Adv. \textbf{2}, 1501772 (2016).

\bibitem{Felinto:2006gw}
D.~Felinto, \textit{et al.}, \enquote{{Conditional control of the quantum states of remote atomic
  memories for quantum networking},} Nature Phys. \textbf{2}, 844 (2006).


\bibitem{Kaneda:2015dn}
F.~Kaneda, \textit{et al.}, \enquote{{Time-Multiplexed Heralded Single-Photon Source},} Optica
  \textbf{2}, 1010 (2015).

\bibitem{Mendoza:2016gr}
G.~J. Mendoza, \textit{et al.}, \enquote{{Active temporal and spatial multiplexing of
  photons},} Optica \textbf{3}, 127 (2016).

\bibitem{Xiong:2016bv}
C.~Xiong, \textit{et al.},
  \enquote{{Active temporal multiplexing of indistinguishable heralded single
  photons},} Nature Commun. \textbf{7}, 10853 (2016).


\bibitem{Lo:2012df}
H.-K. Lo, M.~Curty, and B.~Qi, \enquote{{Measurement-Device-Independent Quantum
  Key Distribution},} Phys. Rev. Lett. \textbf{108}, 130503 (2012).

\bibitem{Lo:2014ex}
H.~K. Lo, M.~Curty, and K.~Tamaki, \enquote{{Secure quantum key distribution},}
  Nature Photon. \textbf{8}, 595 (2014).


\bibitem{Abruzzo:2014it}
S.~Abruzzo, H.~Kampermann, and D.~Bru{\ss},
  \enquote{{Measurement-device-independent quantum key distribution with
  quantum memories},} Phys. Rev. A \textbf{89}, 012301 (2014).

\bibitem{Panayi:2014jr}
C.~Panayi, M.~Razavi, X.~Ma, and N.~L{\"u}tkenhaus, \enquote{{Memory-assisted
  measurement-device-independent quantum key distribution},} New J. Phys. \textbf{16}, 043005 (2014).

\bibitem{Kaneda:2016fh}
F.~Kaneda, K.~Garay-Palmett, A.~B. U'Ren, and P.~G. Kwiat, \enquote{{Heralded
  single-photon source utilizing highly nondegenerate, spectrally factorable
  spontaneous parametric downconversion},} Opt. Express \textbf{24}, 10733
  (2016).

\bibitem{Hong:1987vi}
C.~Hong, Z.~Ou, and L.~Mandel, \enquote{{Measurement of subpicosecond time
  intervals between two photons by interference.}} Phys. Rev. Lett.
  \textbf{59}, 2044 (1987).

\bibitem{Rubenok:2013ip}
A.~Rubenok, J.~A. Slater, P.~Chan, I.~Lucio-Martinez, and W.~Tittel,
  \enquote{{Real-World Two-Photon Interference and Proof-of-Principle Quantum
  Key Distribution Immune to Detector Attacks},} Phys. Rev. Lett.
  \textbf{111}, 130501 (2013).

\bibitem{PhysRevLett.111.130502}
Y.~Liu, \textit{et al.}, \enquote{Experimental measurement-device-independent quantum
  key distribution,} Phys. Rev. Lett. \textbf{111}, 130502 (2013).

\bibitem{PhysRevA.88.052303}
T.~Ferreira~da Silva, \textit{et al.}, \enquote{Proof-of-principle demonstration
  of measurement-device-independent quantum key distribution using polarization
  qubits,} Phys. Rev. A \textbf{88}, 052303 (2013).

\bibitem{PhysRevLett.112.190503}
Z.~Tang, \textit{et al.}, \enquote{Experimental
  demonstration of polarization encoding measurement-device-independent quantum
  key distribution,} Phys. Rev. Lett. \textbf{112}, 190503 (2014).

\bibitem{Comandar:2016em}
L.~C. Comandar, \textit{et al.}, \enquote{{Quantum
  key distribution without detector vulnerabilities using optically seeded
  lasers},} Nature Photon.  \textbf{10}, 312 (2016).

\bibitem{PhysRevLett.117.190501}
H.-L. Yin, \textit{et al.},
  \enquote{Measurement-device-independent quantum key distribution over a 404
  km optical fiber,} Phys. Rev. Lett. \textbf{117}, 190501 (2016).

\bibitem{PhysRevLett.99.180503}
Y.~Adachi, T.~Yamamoto, M.~Koashi, and N.~Imoto, \enquote{Simple and efficient
  quantum key distribution with parametric down-conversion,} Phys. Rev. Lett. \textbf{99}, 180503 (2007).

\bibitem{PhysRevA.75.050305}
W.~Mauerer and C.~Silberhorn, \enquote{Quantum key distribution with passive
  decoy state selection,} Phys. Rev. A \textbf{75}, 050305 (2007).

\bibitem{Kaltenbaek:2006ch}
R.~Kaltenbaek, B.~Blauensteiner, M.~{\.{Z}}ukowski, M.~Aspelmeyer, and
  A.~Zeilinger, \enquote{{Experimental Interference of Independent Photons},}
  Phys. Rev. Lett. \textbf{96}, 240502 (2006).

\bibitem{Fang:2014dk}
B.~Fang, O.~Cohen, M.~Liscidini, J.~E. Sipe, and V.~O. Lorenz, \enquote{{Fast
  and highly resolved capture of the joint spectral density of photon pairs},}
  Optica \textbf{1}, 281 (2014).

\bibitem{Brassard:1993db}
G.~Brassard and L.~Salvail, \enquote{{Secret-Key Reconciliation by Public
  Discussion},} in \enquote{Advances in Cryptology {\textemdash} EUROCRYPT
  {\textquoteright}93,}  (Springer, New York, 1993), p. 410.

\end{thebibliography}

\clearpage
\begin{center}
\section*{Supplementary Information}
\end{center}

\subsection*{Theory of synchronized HSPSs}
Here we show theoretical details of $M$ synchronized HSPSs with imperfect optical components. 
We first define following probabilities: 
\begin{align}
& P_c(k)  = \frac{\mu^k}{(1+\mu)^{k+1}}, 
\label{Pc}
\end{align}
\begin{align}
& P_d(k) = \sum_{l =1 }^k \eta_{d}^l (1-\eta_{d})^{k-l} \binom{k}{l} \left(\frac{1}{D}\right)^{l-1}, 
\label{Pd}
\end{align}
\begin{align}
&P_e(k'|k, j,j') = (T_c T_{QM}^{j-j'+1})^{k'} (1-T_c T_{QM}^{j-j'+1})^{k-k'} \binom{k}{k'}. 
\label{Pt}
\end{align}
$P_c(k)$ is the probability that an SPDC source generates $k$-photon pairs; 
for an SPDC source generating pure heralded photons, its photon number statistics follow a thermal distribution. 
$P_d(k)$ is the probability of a trigger detector click, given that an SPDC source generates a $k$-photon state. 
$\eta_{D}$ is the total transmission of the signal photons from the SPDC crystal to SPDs, and $D$ is the number of SPDs used for a trigger-detector cascade to herald idler photons. 
Here, we assume that the SPDs are ``bucket'' detectors that only discriminate between zero and one-or-more photons, and the detector cascade distributes the signal photons to $D$ detectors with an equal probability ($1/D$). 
$P_e(k' |k,j,j')$ is the idler photon's transmission to an output time slot, where  
$T_{QM}$ denotes the storage efficiency in a QM for a delay time $\tau$, and $T_c$ is the net transmission of the other optics (including an initial delay line, fiber coupling efficiency, etc.). 

We also define $P_h(j)$ as the probability that one HSPS heralds at least one single photon within $j$ time slots: 
\begin{align}
  P_h(j) =& 1-\{1- P_h(1)\}^j,   \label{Ph} \\ 
 P_h(1) =&  \sum_{k =1 }^{\infty}  P_c(k)P_d(k).    \label{Ph1} 
\end{align}
With the above definitions, the probability that all $M$ HSPS's synchronously generate single photons, is given by
\begin{align}
 P_s(M) =& P_l(1|1)^{M} + \sum_{j =2 }^N  \sum_{q =1 }^{M} \binom{M}{q} P_l (1|j)^q P_e(1|j)^{M-q},    \label{Ps(M)}
\end{align}
\begin{align}
P_l(1|j) =&  (1- P_h (j-1) ) \sum_{k' =1 }^{\infty}  P_c(k')P_d(k')P_t(1|k', j,j),  \label{Pl} \\
 P_e(1|j) =&  \sum_{j' =1 }^{j-1} ( 1 - P_h(j-1-j')) \sum_{k' = 1 }^\infty P_c(k')P_d(k') P_t(1| k' ,j,j') ( 1 - P_h(1)) \notag \\
               & + P_h(j-1) \sum_{k' = 1 }^\infty P_c(k')P_d(k') P_t(1| k' ,j,j) \label{Pe}.  
\end{align}
where $N$ is the maximum number of storage time slots. 
$P_l(k|j)$ is the probability that an HSPS initially heralds and generates a $k$-photon state in the $j$-th time slot, 
while $P_e(k|j)$ is the probability of heralding at least one time within $j-1$ time slots and then emitting a $k$-photon state at the $j$-th time slot. 
The first and second term in \eqref{Ps(M)} describe, respectively, the probabilities that all $M$ sources generating single photons in the first time slot and from the second to the $N$-th time slots. 

For $M= 2$, as used for our experiment, we extended Eq. \eqref{Ps(M)} to calculate multi-photon emission probabilities from each HSPS. 
The probability of Source A and B respectively producing $k_A$ and $k_B$ photons is given by
\begin{align}
 P'_s(k_A,k_B) =& P_l(k_A|1)P_l(k_B|1) + \sum_{j =2 }^N   (P_l (k_A|j) P_e (k_B|j) + P_l (k_B|j) P_e (k_A|j)+ P_l (k_A|j) P_l (k_B|j) ).     \label{Ps'}
\end{align}
The experimental parameters used for our theoretical estimations are shown in Table \ref{parameters}. 
\begin{center}
\begin{table}[t!]
\caption{Experimental parameters  }
\begin{tabular}{   lcc }      
      
   Description                  &   Symbol                               &   Value                 \\  \hline 
  Mean photon number per pulse    & $\mu$                                       & 0.013   \\
  Trigger-mode system detection efficiency    & $\eta_t$                                       & 0.18  \\
 Optical channel transmission   &$T_c$                                       & 0.083                     \\
  Quantum memory transmission    & $T_{QM}$                                      & 0.988 \\  
  HOMI/BSM detector efficiency    & $\eta_d$                                       & 0.75                        \\ \hline
\label{parameters}
\end{tabular}

\end{table}
\end{center}

\subsection*{Spectral characterization of HSPSs}
\begin{tiny}
\begin{figure}[t!]
  \includegraphics[width=0.7\columnwidth,clip]{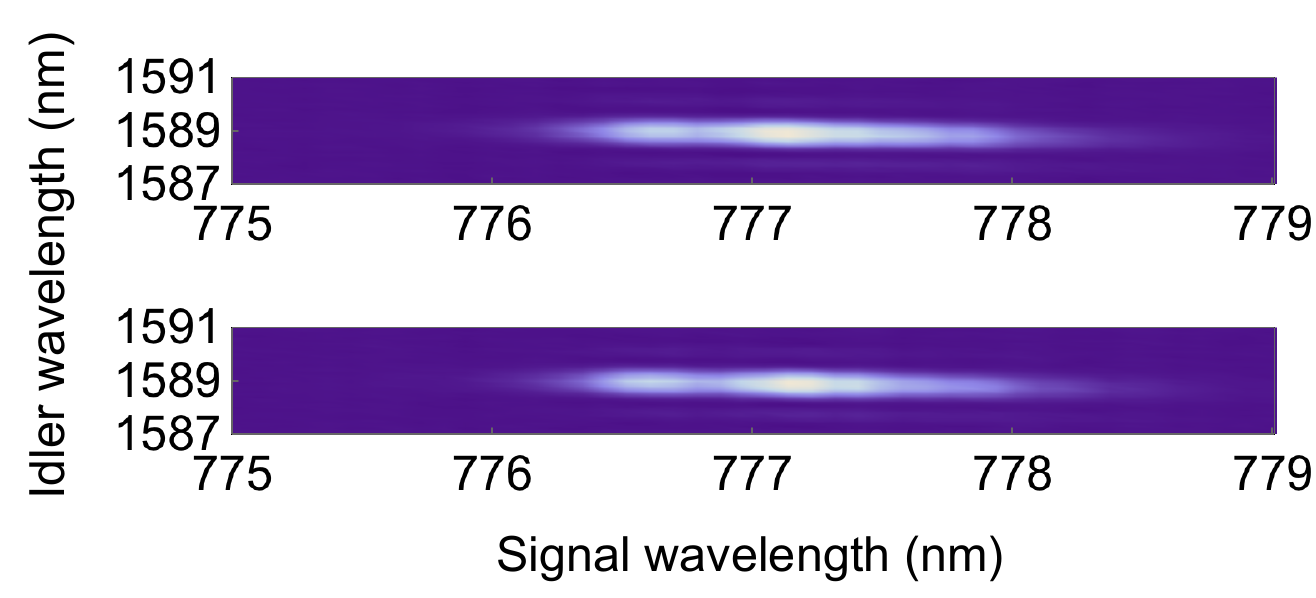}
   \caption{Observed JSI for two SPDC sources.  }
\label{JSI}
\end{figure}
\end{tiny}

In order to estimate an attainable visibility of our HOMI measurement, we measured joint spectral intensities (JSIs) of the two SPDC sources (see Fig. \ref{JSI}), using frequency-resolved optical parametric amplification \cite{Fang:2014dk,Kaneda:2016fh}.
As shown in Fig. \ref{JSI}, the SPDC sources have a very similar JSI, each of which is estimated to generate heralded single-photons with 97.1\% purity,  assuming for no spectral phase shift in the JSI. 
We then estimate the final indistinguishability of the heralded photons from the two independent sources to be 96.4\%; our experimental visibility ($95.7 \pm 1.5 $\%) is very close to this estimate.

\subsection*{Secure key rate evaluation and possible improvements in MDI-QKD}
We determined the gain and quantum bit error rate (QBER) respectively as $Q_W = (C_{00}+C_{11}+C_{01}+C_{10})/N_p$, $e_{W} = (C_{ 00}+C_{11})/(C_{ 00}+C_{11}+C_{01}+C_{10})$, where $W \in [Z =\{ \ket{e}, \ket{l} \} , X= \{ \ket{+}, \ket{-} \} ]$ is the basis choice, and $C_{ij}$ ($i,j \in \{0,1\}$) is the number of coincidence counts for input qubits $\ket{ij}$, given a total number of pump pulses $N_p$~\cite{Lo:2012df}. 
In our proof-of-principle implementation, a lower bound of secure key rate $R$ is estimated by assuming that our HSPS is an ideal single-photon source (i.e., ignoring the low multi-pair contributions):
\begin{align}
R = Q_{Z}^{L}\left\{1-h(e_{X}^{U})-f_{e}h(e_{Z}^{U})\right\}.
\end{align}
Here, $Q_{Z}^{L}$, $e_{X}^{U}$, $e_{Z}^{U}$ are the lower (L) and upper (U) bounds of gain and QBER due to statistical fluctuations (we consider 3 standard deviations); $h(x) = -x \log x -(1-x) \log (1-x)$ is the binary entropy, and $f_{e}= 1.16$ is the error correction inefficiency factor \cite{Brassard:1993db}. Our observed QKD parameters as well as estimated secure key rates $R$ are shown in Table~\ref{QKDparameters}.

\begin{table}[t!]
\caption{Experimental MDI-QKD quantities and estimated key rate. }
\begin{tabular}{   lcc }      \hline
                                          &   With synchronization                               &   Without synchronization                 \\  \hline 
  $Q_Z$ ($\times 10^{-7}$ per pulse)    & $1.976   \pm   0.033$                & $0.0688   \pm   0.0060$  \\
  $Q_X$ ($\times 10^{-7}$ per pulse)     &$1.969   \pm   0.033$                   & $0.0718   \pm   0.0062$                   \\
  $e_Z$     & $0.0771 \pm   0.0045$                & $0.0747 \pm   0.0237$  \\
  $e_X$     & $0.0797 \pm   0.0046$                &$0.0791 \pm   0.0239$                     \\
  $N_p$       & $4\times 10^9$ & $4\times 10^9$   \\
  $R $  ($\times 10^{-7}$ per pulse)  & 0.212                                     & -0.00107   \\ \hline
\label{QKDparameters}
\end{tabular}
\end{table}

We expect that our current observed secure key rate can be increased by a factor of $\sim 500$ by using efficient optics, a deterministic time-bin encoding method, and decoy-state method with parameter optimizations. 
Since we used this postselective method for simplicity of implementation (see Methods), each time-bin encoder has only 22\% transmission; however, lossless and deterministic encoding can be achieved, for example, by using time-bin entangled photon sources instead of an HSPS, detecting trigger photons with projection onto corresponding time-bin states. 
$e_Z$ can be reduced to $< 1$\% by extending the time bin's separation from our current 25 ps to $\geq 200$ ps such that SNSPDs can resolve time-bin states. 
This has already been demonstrated in previous weak-coherent-pulse(WCP)-based MDI-QKD experiments \cite{Rubenok:2013ip,PhysRevLett.111.130502,PhysRevLett.117.190501}.  
Our BSM setup has only $\sim 60$\% transmission due to the high fiber-coupling loss; 
employing free-space BSM would make this loss negligibly small. 
Overall, these improvement would increase $Q_z$ and $Q_x$ by a factor of $(0.22\times 0.6)^{-2} = 57$, reduce $h(e_z)$ from 0.408 to  $<0.081$, and therefore $R$ increased by a factor of $\sim250$. 
In addition, decoy-state methods together parameter optimizations can substantially increase the secure key rate, potentially by as much as another factor of $\sim$ 10--20.
The resulting $R$ could then be $\sim 10^{-5}$ bit per pulse, which is comparable to WCP-based MDI-QKD experiments \cite{Rubenok:2013ip,PhysRevLett.111.130502,PhysRevA.88.052303,PhysRevLett.112.190503,Comandar:2016em,PhysRevLett.117.190501}).
Finally, a faster Pockels cell which can repeat a synchronization process immediately after previous one, is able to increase the success event rate of the BSM as well as the secure key rate \textit{per second} by a factor of 2.5.

\end{document}